\documentclass{iopjournal}

\usepackage{fancyhdr}

\fancypagestyle{arxiv}{
    \fancyhf{}
    \fancyfoot[C]{\thepage}
    \renewcommand{\headrulewidth}{0pt}
    \renewcommand{\footrulewidth}{0pt}
}

\usepackage[utf8]{inputenc}
\usepackage[T1]{fontenc}
\usepackage[english]{babel}
\usepackage{amsmath}
\usepackage{amssymb}
\usepackage{mathtools}
\usepackage{bbold}
\usepackage{graphicx}
\usepackage{xcolor}
\usepackage[numbers]{natbib}
\usepackage{hyperref}
\hypersetup{
    colorlinks=true,
    citecolor=teal,
    linkcolor=teal,
    urlcolor=teal
}

\usepackage[utf8]{inputenc}
\usepackage[english]{babel}
\usepackage[T1]{fontenc}
\usepackage{hyperref}

\usepackage{bbold}
\usepackage{mathtools}
\usepackage{graphicx} 
\usepackage{xcolor}

\renewcommand{\Re}{\mathop{\text{Re}}\nolimits}
\renewcommand{\Im}{\mathop{\text{Im}}\nolimits}

\def\be{\begin{equation}}
\def\ee{\end{equation}}

\newcommand{\beq}{\begin{equation}}
\newcommand{\eeq}{\end{equation}}

\newcommand*{\C}{\text{C}}

\renewcommand*{\L}{\text{L}}
\newcommand*{\R}{\text{R}}

\newcommand{\kBT}{k_\text{B}T}

\begin{document}

\title{Capturing an Electron in the Virtual State}

\author{Alok Nath Singh$^{1,2,*}$,
Bibek Bhandari$^2$,
Rafael S\'anchez$^{3,4,5}$,
and Andrew N. Jordan$^{2,1,6}$}

\affil{$^1$Department of Physics and Astronomy, University of Rochester,
Rochester, New York 14627, USA}

\affil{$^2$Institute for Quantum Studies, Chapman University,
Orange, California 92866, USA}

\affil{$^3$Departamento de F\'isica Te\'orica de la Materia Condensada,
Universidad Aut\'onoma de Madrid, 28049 Madrid, Spain}

\affil{$^4$Condensed Matter Physics Center (IFIMAC),
Universidad Aut\'onoma de Madrid, 28049 Madrid, Spain}

\affil{$^5$Instituto Nicol\'as Cabrera (INC),
Universidad Aut\'onoma de Madrid, 28049 Madrid, Spain}

\affil{$^6$The Kennedy Chair in Physics, Chapman University,
Orange, California 92866, USA}

\affil{$^*$ Author to whom any correspondence should be addressed.}

\email{aloknsingh28@gmail.com}

\begin{abstract}
We address a foundational question in quantum mechanics: Can a particle be
directly found in a classically forbidden virtual state? We instantiate this
conceptual question by investigating the traversal of electrons through a
tunnel barrier, which we define in a triple quantum dot (TQD) system where the
occupation of the central dot is energetically avoided.
The motivation behind this setup is to determine whether the central dot is
occupied during a virtual transition when it is explicitly monitored. We
investigate this problem in two different limits of continuous measurement:
stochastic quantum diffusion and quantum jumps. We find that, even though
individual trajectories differ considerably across these limits, measurement
leads to a higher average occupation of the central dot. We also find that a
suitably strong measurement can significantly boost the tunneling current—an
effect we call \textit{measurement-assisted tunneling}.
Our results demonstrate that observation fundamentally reshapes tunneling
dynamics, resolving the apparent paradox of detecting a particle in a
classically forbidden region: weak measurements partially localize the
particle, whereas strong measurements enforce a discontinuous either-or
outcome of detection or no detection.
\end{abstract}

\thispagestyle{arxiv}

\section{Introduction}
\label{sec:intro}
Quantum tunneling is both a paradigmatic manifestation of the particle/wave aspect of quantum physics, and a deeply puzzling effect from a foundations of physics perspective. 
Much debate continues about simple questions, such as ``How long does a particle remain in the classically forbidden region?'' \cite{buttiker_traversal_1982,landauer1994,Steinberg1995,shafir2012, choi_operational_2013, ramos2020, sinclair2022, thompson2025}. 
It is well known that measurements of quantum properties yield results that are classically well defined - a particle will be found in a particular position if that is indeed what is being measured.  Here we examine the question of measurements on tunneling particles.  If we interrogate {\it where} a tunneling particle is found as it passes from one side to another of a classically forbidden region, a dichotomy arises:  On one hand, if we assume a continuous trajectory - that to pass from point A to point C, the particle must go through point B (having a higher energy than A and C), then a measurement of the particle's position at point B must reveal itself.  In that case,  the particle must have sufficient energy to be allowed at the position.  Since the particle was assumed to not have a sufficient amount of energy, where did it come from?  On the other hand, an opposing possibility is that by the act of looking at position B we simply prevent the tunneling from happening in the first place---the measurement spoils the effect.


{Properly addressing these questions requires an explicit model of the monitored device and the way it is coupled to the detector.}
Crucially, how energy is exchanged between a measured quantum system and its detector---and whether a measurement can supply, absorb or merely redistribute energy within the system---remains subtle and is an active topic of theoretical and experimental study \cite{manikandan2022efficiently, linpeng2024quantum, dassonneville2026directly}. By explicitly monitoring a tunnel barrier, our setup places these energy-exchange processes front and center.
{
We emphasize that our goal is not to establish the presence of an electron in the virtual state during an unmonitored cotunneling process. Rather, we investigate how continuous measurement modifies the tunneling dynamics and under what conditions measurement backaction can induce a measurable population of the intermediate state.} In this setting, the detector both reads out charge and modifies the energetics of the tunneling process, allowing us to predict how measurement-mediated energy transfers can facilitate or inhibit barrier traversal.

\begin{figure}[!htb]
\centering
\fbox{%
\begin{minipage}{\columnwidth}
\centering
\includegraphics[width=.6\linewidth]{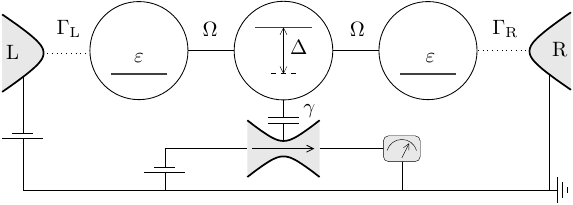}
\caption{\label{fig:scheme} Scheme of the triple quantum dot coupled to two terminals ($l={\rm L,R}$) via tunneling rates $\Gamma_l$. The energies of the singly-occupied quantum dots are represented: the central dot is split by $\Delta$ with respect to the other two, at $\varepsilon$. The nearest neighbor hopping is $\Omega$. The charge of the central dot is monitored by a coupled QPC measuring with a rate $\gamma$. 
}
\end{minipage}}

\end{figure}  

{In the results presented here, we will show that depending on {\it how} the measurement is done, both of the above situations can be manifested.}  We tune the strength of the measurement from arbitrarily weak to projective.  We consider both diffusive and jumpy continuous measurements and discuss how we can infer the behavior of the tunneling particle from the continuous measurement results.  
To simplify the conceptual presentation above, we consider a line of three tunnel coupled quantum dots where the central dot is energetically higher than the other dots, as sketched in Fig.~\ref{fig:scheme}.  In such a situation, a tunnel current induced by an electrical bias across the chain is only possible if an electron transits through the virtual state in the central dot---which plays the role of the classically forbidden potential barrier.  
The position measurement is realistically made via the charge of the electron.  By capacitively coupling the central dot to a nearby electrical conductor that is also electrically biased, we can make real time measurements of the occupation of the central dot in a way whose measurement strength can be tuned from no measurement at all, to a rapid projective measurement of the dot's occupation. The electrical conductor can function in the diffusive quantum limit when operating in the weakly responding regime, or the quantum jump limit when operating in the tunneling regime and counting transmitted electrons that are only permitted to transport when the central dot is empty.
{ These two realizations result in qualitatively distinct quantum trajectories, but we find that the ensemble-average dynamics are the same given the same measurement strength.}
While the basic physics here can be realized in many different analogous physical set-ups, this quantum dot realization is both experimentally realistic, and conceptually clear and satisfying.


Related works have explored the lifetime and applications of virtual states in quantum dot-based setups. In Refs. \cite{romito_weak_2014, zilberberg2014}, the authors examined the lifetime of a virtual state through weak measurements and post-selection based on a classically forbidden cotunneling process in a single quantum dot \cite{deFranceschi_electron_2001,yamada_variation_2003,schleser_cotunnelling_2005}. Investigations of virtual states in the context of quantum transport properties have also been conducted on double \cite{gustavsson2008} and triple \cite{zilberberg2019} quantum dot setups. Additionally, virtual states were employed in triple quantum dot (TQD) based setups to demonstrate cotunneling between the outer quantum dots \cite{amaha_resonance_2012,busl_bipolar_2013,braakman_long_2013,sanchez_longrange_2014}, showing promise for application in quantum computing.
{ 
More broadly, measurement backaction in mesoscopic transport is related to a wider class of dephasing-controlled transport phenomena. In ``which-path’’ interference experiments, a nearby QPC acquires path information about an electron and suppresses coherent transport through measurement-induced dephasing \cite{buks_dephasing_1998,aleiner_dephasing_1997}. Similar dephasing mechanisms have been studied in continuously monitored quantum-dot systems and mesoscopic conductors \cite{gurvitz1997measurements,gurvitz_microscopic_1996,clerk2003quantum}. Moreover, under suitable conditions, dephasing can also enhance transport by mitigating coherent bottlenecks, an effect discussed in the context of noise-assisted and environment-assisted transport \cite{brandes2005coherent,plenio2008dephasing,contreras2014dephasing,rebentrost2009environment}. Taken together, these earlier studies separately established (i) the role of virtual states in cotunneling, (ii) the influence of QPC backaction on coherent transport, and (iii) the enhancement of transport under moderate dephasing. The present work brings these ingredients together in a single experimentally motivated setting, where a continuously monitored detector is coupled directly to a detuned virtual state, allowing us to connect measurement-induced occupation, transport enhancement, conditioned quantum trajectories, and transport--detector current correlations within one unified framework.}


By tracking the stochastic occupation trajectories of the monitored virtual state, this work explores how continuous measurement modifies the dynamics of virtual-state transport. We find that in both the diffusive and quantum-jump measurement limits, detector backaction substantially increases the average occupation of the detuned central dot compared with the unmonitored case. This enhanced localization is accompanied by an increase in the average tunneling current at intermediate measurement strengths, whereas sufficiently strong monitoring suppresses transport through the quantum Zeno effect.

\section{Model}
\label{sec:model}
We consider a TQD system attached to two electronic reservoirs as shown in Fig.~\ref{fig:scheme}. The central dot is continuously monitored by detecting the current through a capacitively-coupled QPC \cite{gustavsson:2006,fujisawa:2006}. The Hamiltonian for the TQD system is given by
\begin{align}
\label{eq:Ham}
\hat{H}_{\text{TQD}} &= \sum_{i}\varepsilon_{i}\hat{c}^{\dagger}_{i}
\hat{c}_{i}^{}-\sum_{i\neq j}\left(\Omega_{ij}\hat{c}^{\dagger}_{i}\hat{c}_{j}^{}+\text{H.c.}\right),
\end{align}
where $\hat{c}_i$ are the annihilation operators for an electron in the quantum dot, $i=\{{\rm L, C, R}\}$ with energy $\epsilon_i$. $\hat n_i = \hat c_i^\dagger c_i$ is the number operator for the quantum dot $i$ and $\Omega_{ij}$ gives the tunnel coupling strength between quantum dots $i$ and $j$. We consider symmetric nearest neighbor coupling such that, $\Omega_{\rm LC} = \Omega_{\rm RC} = \Omega$ whereas $\Omega_{\rm LR} = 0 $. Additionally, we assume strong onsite and inter-dot Coulomb interaction, much larger than any other energy scale, resulting into a spinless interaction-free Hamiltonian of Eq.~(\ref{eq:Ham}). 
The reservoir Hamiltonian can be expressed as, 
\begin{equation}
\hat H_{\rm res} = \sum_{l={\rm L,R}}\sum_k\epsilon_{k,l}\hat{d}_{k,l}^\dagger \hat d_{k,l},
\end{equation}
where $\hat{d}_{k,l}$ are the annihilation operators for an electron with energy $\epsilon_{k,l}$ in bath $l={\rm L, R}$. The TQD system-reservoir coupling is given by 
\begin{equation}
\hat{H}_{\rm tun} = \sum_{l,k}\gamma_{l}\hat{d}^{\dagger}_{k,l}
\hat{c}_{l}+\text{H.c.},
\end{equation}
where $\gamma_l$ gives the system-reservoir coupling strength.

We consider a configuration where $\varepsilon_{\rm L} = \varepsilon_{\rm R} = \varepsilon$ and $\varepsilon_{\rm C} = \varepsilon + \Delta$, allowing electrons to resonantly transfer from L to R \cite{ratner_bridge_1990}. When $\Delta \gg \Omega$, the central dot does not hybridize with the other two, thereby avoiding its occupation.  
{ However, in this weakly hybridized limit, transport still occurs via second‐order cotunneling---direct, virtual transitions between the outer dots through the detuned central dot \cite{busl_bipolar_2013,braakman_long_2013,sanchez_longrange_2014,amaha_resonance_2012}. An electron from L can tunnel into C for a short period of time consistent with the time-energy uncertainty  $\Delta E \Delta t \le \hbar$, after which it either tunnels back to L or tunnels forward to R \cite{nazarov_quantum_2009}. There’s no net energy violation once the full second-order process completes.}
Thus, the TQD serves as a discrete analog of a tunnel barrier, with the detuning of the central dot setting the barrier height. A perturbative expansion yields the effective coupling for virtual tunneling as $\Omega_{\rm eff} = \Omega^2 / \Delta$ \cite{superexchange}.
{ For the detector to be sensitive to transient occupation of the virtual state, its characteristic response time must be shorter than the intrinsic virtual-state lifetime set by the detuning, $\hbar\Delta^{-1}$.}

\section{Diffusive quantum measurement}
\label{sec:diffusive}
We first look at the situation where we make a weak continuous measurement of the occupation of C. Inside the QPC, transmittance of the saddle point constriction depends on whether an electron in C is present~(with transmittance $T_1$) or not~(with transmittance $T_0$). In the diffusive quantum measurement limit, transmittance is not affected significantly by the occupation, i.e., $|T_1-T_0|\ll (T_1+T_0)/2$ \cite{andrew_book}.

{Under the Born--Markov approximation for the source and drain reservoirs, the conditional state of the TQD can be described by a stochastic master equation in the It\^o representation \cite{andrew_book,gisin1992quantum}. The TQD dynamics are then governed by}
\beq\label{eq:ME}
\frac{d\hat\rho}{dt}=-\frac{i}{\hbar}[\hat{H}_{\text{TQD}},\hat\rho]+\sum_{l}({\cal L}_{\Gamma,l+}\hat\rho+{\cal L}_{\Gamma,l-}\hat\rho)+{\cal L}_\gamma\hat\rho,
\eeq
where ${\cal L}_{\Gamma,l\pm}$ denote the usual Lindblad superoperators corresponding to the reservoirs: 
\begin{equation}
{\cal L}_\lambda=\hat{L}_\lambda^{}\hat\rho\hat{L}_\lambda^{\dagger}-\frac{1}{2}\left(\hat{L}_\lambda^{\dagger}\hat{L}_\lambda^{}\hat\rho+\hat\rho\hat{L}_\lambda^{\dagger}\hat{L}_\lambda^{}\right)
\end{equation}
with the jump operators related to electron tunneling in, $\hat{L}_{l+}=\sqrt{\Gamma_{l+}}|l\rangle\langle0|$, or tunneling out, $\hat{L}_{l-}=\sqrt{\Gamma_{l-}}|0\rangle\langle l|$, of quantum dot $l={\rm L,R}$.
The tunneling rates are given by $\Gamma_{l+}=\Gamma_l f(\varepsilon_l-\mu_l)$ and $\Gamma_{l-}=\Gamma_l[1-f(\varepsilon_l-\mu_l)]$, where $f(E)=[1+\exp{(E/\kBT)}]^{-1}$ is the Fermi function and $\mu_l$ is the electrochemical potential of reservoir $l$. $\Gamma_l=2\pi\hbar^{-1}|\gamma_l|^2\nu_l$, with $\nu_l$ being the density of states in lead $l$, is the tunneling rate between the dot $l$ and reservoir $l$. 
{ The conditional evolution of the system is described by the measurement superoperator written in the Itô representation as
\beq \label{eq:Lindbladian_measurement}
\begin{aligned}
{\cal L}_\gamma \hat{\rho}=~&\hat{L}_\gamma^{}\hat\rho\hat{L}_\gamma^{\dagger}-\frac{1}{2}\left(\hat{L}_\gamma^{\dagger}\hat{L}_\gamma^{}\hat\rho+\hat\rho\hat{L}_\gamma^{\dagger}\hat{L}_\gamma^{}\right)
+ (\hat{L}_\gamma\hat\rho + \hat\rho\hat{L}_\gamma^{\dagger} - \langle \hat{L}_\gamma + \hat{L}_\gamma^\dagger \rangle \hat\rho) \frac{dW}{dt}
\end{aligned}
\eeq
where the measurement jump operator is $\hat{L}_\gamma = \sqrt{\gamma}|C\rangle \langle C|$. The stochastic term describes the measurement backaction associated with a continuous detector record. Here $dW(t)$ is a real Wiener increment in the Itô calculus with zero mean and variance $dt$, satisfying $dW(t)^2=dt$,
with all higher-order products vanishing to leading order in $dt$. The Itô form is particularly convenient for numerical trajectory simulations, since the stochastic increment at time $t$ is statistically independent of the system state at later times \cite{gisin1992quantum}.} 
Measurement strength is given by $\gamma=\frac{e^4V^2}{2h^2S_I}(T_0-T_1)^2$ \cite{andrew_book}, where $V$ is the voltage applied across the QPC, $h$ is Planck's constant, $S_I$ is the shot noise associated with the QPC current, and $T_0$ ($T_1$) are the transmission probabilities of the QPC when the central dot is unoccupied (occupied). {We take $\hbar=1$ in the rest of the text.}
\

\begin{figure}[!htb]
\centering
\fbox{%
\begin{minipage}{\columnwidth}
\centering
\includegraphics[width=.9\linewidth]{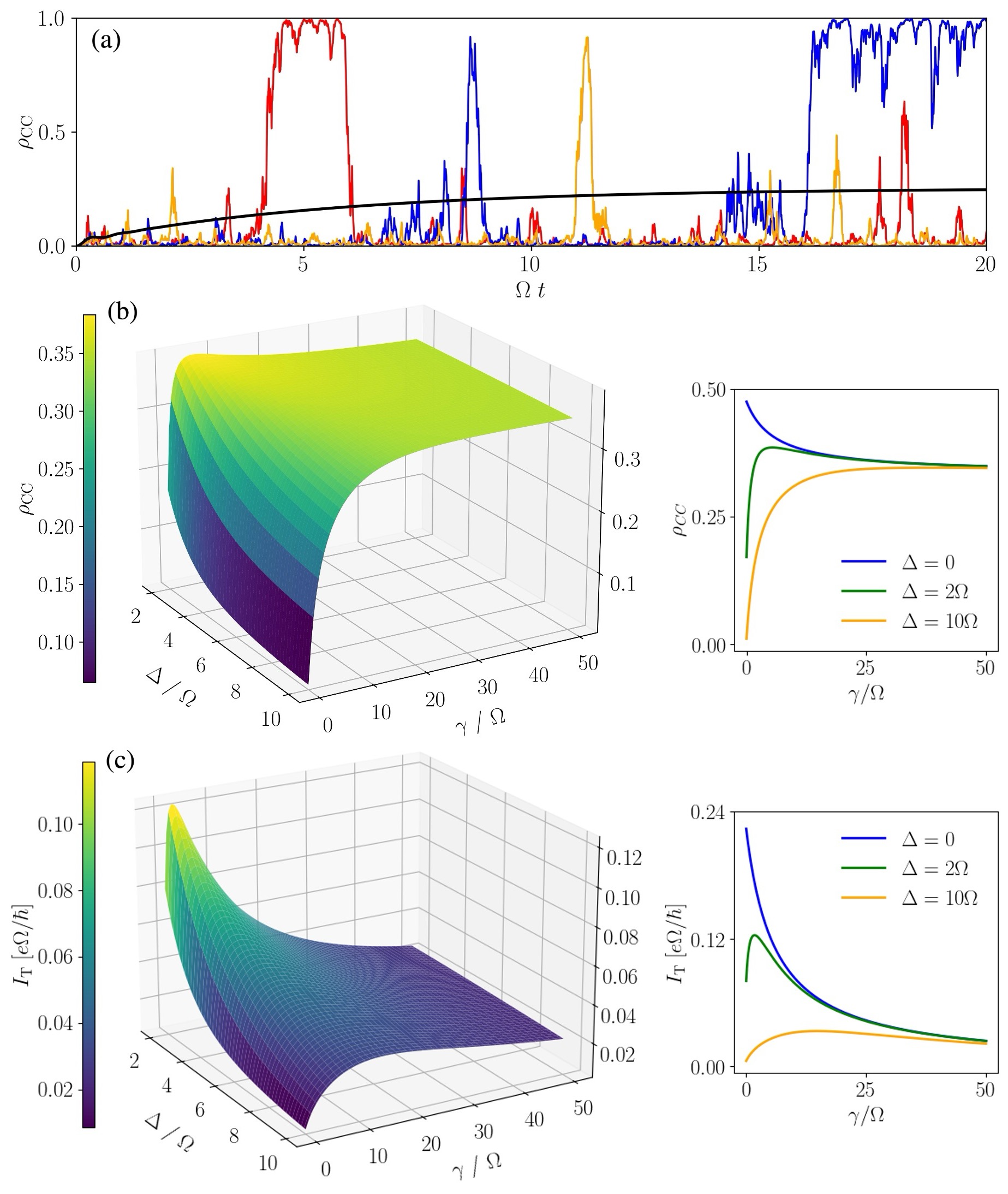}
\caption{ 
{(a)} Stochastic evolution of the central-dot occupation as a function of time. {The black curve shows the ensemble-averaged evolution obtained by averaging over many measurement realizations. In contrast, the three colored curves show individual conditioned quantum trajectories, each corresponding to a different sequence of measurement outcomes generated by the detector. For visual clarity, the single-trajectory results are smoothed using a rectangular time-averaging window of width $0.1~\Omega^{-1}$.} Initial state is taken to be ${\rm |L\rangle\langle L|}$, { for $\Delta=10~\Omega,$ $\gamma=10~\Omega$, and $dt=10^{-4}~\Omega^{-1}$. {(b, c)}  Steady state, ensemble averaged central dot occupation ($\rho_{\rm CC}$) and current through the TQD ($I_{\rm T}$) tuning $\Delta$ and $\gamma$. { Side panels show the same for a select values of $\Delta$, including $\Delta=0$ which is not shown in the surface plot. }Other parameters common to (a, b, c): $\Gamma_{\rm L}=10~\Omega \text{, and } \Gamma_{\rm R}=8~\Omega$.} }
    \label{fig:diffusive}
\end{minipage}}
\end{figure} 

We consider the situation where the reservoirs are at the same temperature, but have a large voltage bias across them, such that $f(\varepsilon_{\rm L}-\mu_{\rm L})\approx1$ and $f(\varepsilon_{\rm R}-\mu_{\rm R})\approx0$. 
{ In this regime, electron transport through the TQD is effectively unidirectional. 
We further assume the wide-band approximation for the source and drain reservoirs, such that the tunneling rates $\Gamma_{\rm L}$ and $\Gamma_{\rm R}$ are energy independent. 
Under these assumptions, the reservoir correlation times are much shorter than the intrinsic system timescales, justifying the Born--Markov treatment of the lead couplings. 
Moreover, in the regime $\Omega \ll \Gamma_{\rm L},\Gamma_{\rm R}$ considered throughout this work, the resulting local master equation provides an accurate description of the transport dynamics \cite{gurvitz_microscopic_1996,harbola2006quantum,potts:2021}. As long as these conditions are met and the local master equation holds, our analysis is robust against changes in the bath couplings, as verified by numerical simulations. }
Note that because of this large bias, our results do not depend on $\varepsilon$. 

We use Eq.~(\ref{eq:ME}) to find a coupled differential equation for each element of $\hat{\rho}$ (see Appendix~\ref{app:SME} for complete master equations), which are then solved numerically to arrive at the state evolution. 
Figures \ref{fig:diffusive}(a) and \ref{fig:diffusive}(b) respectively show the stochastic evolution of $\rho_{\rm CC}$ over time, and the steady-state, ensemble-averaged result for the central dot occupation, $\rho_{\rm CC}$. 
While undergoing continuous monitoring, it quickly rises up and then saturates to $\rho_{\rm CC}\to1/3$, with increasing measurement strength. {In the same regime, we find $\rho_{\rm LL}\to2/3$ and $\rho_{00},\rho_{\rm RR}\to 0$.} An explanation as to why $\rho_{\rm CC}$ reaches this value is provided later in Section~\ref{sec:largegamma}.
Note that while the ensemble average value of $\rho_{\rm CC}$ saturates to a finite value smaller than 1, individual trajectories can occasionally reach unit probability. 
{ Note {as well} that the energy needed to populate C is provided by the measurement device, since the measurement operator, $\hat{L}_\gamma$, does not commute with $\hat {H}_{\rm TQD}$, cf. Eq.~\eqref{eq:Ham} \cite{bhandari2020continuous}}.
{The relative value of $\rho_{CC}$ and $\rho_{RR}$ (and therefore the current) gives some intuition of the effect of the measurement. In the absence of the detector, the ratio $r=\rho_{CC}/\rho_{RR}=(\Gamma_{\rm R}^2+4\Omega^2)/4\Omega^2$ is independent of the barrier height, $\Delta$, suggesting that transport through the device cannot avoid a certain population of the barrier state. In the high barrier regime ($\Delta\gg\Omega$), both are similarly suppressed as $\Omega^2/\Delta^2$. See the unmonitored steady-state occupations in Appendix~\ref{app:nomeasurement} for further details. The presence of the detector enhances the ratio $r$, i.e., $r(\gamma\neq0)>r(\gamma=0)$, even in the measurement-induced tunneling regime, with $r$ diverging in the strong measurement regime: measurement localizes the electron in C, preventing it from tunneling to R. }

The current through the TQD, $I_{\rm T}=e\Gamma_{\rm R}\rho_{\rm RR}$, shown in Fig.~\ref{fig:diffusive}(c), exhibits a richer behavior. For large $\Delta$, $I_{\rm T}$ is nearly zero in the absence of measurement, as electrons remain localized in the left dot. As the measurement strength increases, backaction populates dot C, enabling transport into dot R and enhancing $I_{\rm T}$. We refer to this effect as {\it measurement-assisted tunneling}. 
{Under steady-state conditions, the current is proportional to the interdot coherences, 
$I_{\rm T}=2e\Omega\,\Im\rho_{\rm LC}=2e\Omega\,\Im\rho_{\rm CR}$, indicating that intermediate measurement strengths enhance coherence between neighboring dots. 
At the ensemble level, the resulting current enhancement is closely related to previously studied dephasing-assisted transport, where moderate coupling enhances transport before strong coupling leads to suppression \cite{contreras2014dephasing}. 
In this sense, measurement-assisted tunneling may be viewed as a monitored realization of the same general mechanism. 
The key distinction in the present work is that the dephasing channel originates from an explicitly monitored detector, which also injects energy into the TQD and provides access to conditioned quantum trajectories and experimentally measurable transport--detector current correlations.
}

In the strong measurement regime, irrespective of the detuning, we observe the quantum Zeno effect \cite{facchi2008quantum, greenfield2025unified}: continuous monitoring freezes the movement of the electron for extended periods of time, thus bringing the average current down toward zero. 
With these ensemble-averaged results we can also talk about how long, on average, an electron {\it dwells} inside the central dot (see Appendix~\ref{app:dwell} for more information).

{It is interesting to compare the large $\Delta$ regime with the hopping limit ($\Delta=0$) in which the tunnel barrier is absent. Analytically:
\begin{gather}
\begin{aligned}
\label{eq:hoppinglimit}
\rho_{\rm CC}(\Delta=0)&=\frac{\Gamma_{\rm L}(\gamma\Gamma_{\rm R}+\Gamma_{\rm R}^2+4\Omega^2)}{\Gamma_{\rm L}\Gamma_{\rm R}(3\gamma+2\Gamma_{\rm R})+4\Omega^2(\Gamma_{\rm R}+3\Gamma_{\rm L})}, \text{ and}\\
I_{\rm T}(\Delta=0)&=\frac{4\Gamma_{\rm L}\Gamma_{\rm R}\Omega^2}{\Gamma_{\rm L}\Gamma_{\rm R}(3\gamma+2\Gamma_{\rm R})+4\Omega^2(\Gamma_{\rm R}+3\Gamma_{\rm L})}.
\end{aligned}
\end{gather}
Indeed, here the measurement has an opposite effect on the transport: both $\rho_{CC}$ and $I_{\rm T}$ decrease monotonically with the measurement strength, as shown in Fig.~\ref{fig:diffusive}~(b, c) side panels respectively. The measurement introduces decoherence between left and center dots which tends to trap the electron in L, in this situation, leading to a {\it measurement-suppressed hopping} effect.}

\subsection{Strong measurement limit}
\label{sec:largegamma}
{ The large $\gamma$ limit, where coherences vanish due to the measurement, can be understood as follows. The quantum Zeno effect restricts the motion of the electron by localizing it at a particular site, either L or C, for long periods of time. The occupations in R and the empty state are avoided due to the high bias.
We get compact closed-form analytic expressions, applicable for all $\Delta$, by neglecting the coherence between the disconnected outer dots, $\rho_{\rm LR}$, in Eq.~\eqref{eq:ME}: 
\begin{gather}
\begin{aligned}
I_{\rm T}(\gamma\to\infty)&=\bigg[\frac{1}{\Gamma_{\rm L}}+\frac{3}{\Gamma_{\rm R}}+\frac{1}{\Omega^2}\!\left(\frac{\Delta^2}{\Gamma_{\rm R}+\gamma/2}+\Gamma_{\rm R}+\gamma/2\right)
+\frac{1}{\Omega^2}\!\left(\frac{\Delta^2}{\gamma}+\frac{\gamma}{4}\right)\bigg]^{-1}, \text{ and}\\
\rho_{\rm CC}(\gamma\to\infty)&=\left[\frac{1}{\Gamma_{\rm R}}+\frac{1}{2\Omega^2}\!\left(\frac{\Delta^2}{\Gamma_{\rm R}+\gamma/2}+\Gamma_{\rm R}+\gamma/2\right)\right] I_{\rm T}(\gamma\to\infty),
\end{aligned}
\end{gather}
which approach the exact solutions in the projective-measurement limit. In that limit, where $\gamma$ is the largest energy scale, the current is suppressed as $I_{\rm T}\propto\rho_{\rm RR}\sim4\Omega^2/3\gamma$ and $\rho_{\rm CC}$ approaches 1/3. In the absence of coherence, transport is governed by classical rate equations, with the current proportional to the population imbalance between adjacent dots \cite{gurvitz1997measurements, gurvitz_microscopic_1996, facchi2008quantum}. At steady state, this current needs to be the same across the system, requiring $\rho_{\rm LL}-\rho_{\rm CC} = \rho_{\rm CC}-\rho_{\rm RR}$. Imposing $\rho_{\rm RR}\approx\rho_{00}\approx0$ implies $\rho_{\rm LL}\approx2/3$ and $\rho_{\rm CC}\approx1/3$. 
The fact that $\rho_{LL}\approx2\rho_{CC}$ is hence dictated by the details of the measurement and the transport configuration rather than by the existence of the tunnel barrier. { We verified numerically that this limit is indeed robust against changes in $\Delta$ and $\Omega$ as long as $\gamma$ is much larger than both.}
}

\begin{figure}[]
\centering
\fbox{%
\begin{minipage}{\columnwidth}
\centering
\includegraphics[width=1\linewidth]{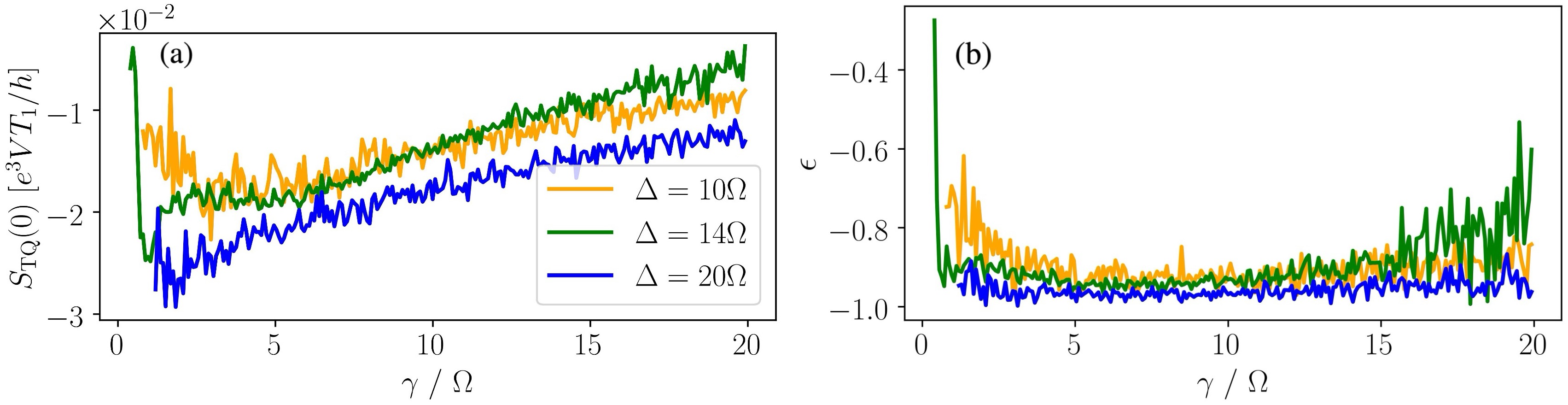}
\caption{Zero frequency cross correlation \textbf{(a)}, and Pearson coefficient \textbf{(b)} between the current through the TQD and the detector as a function of measurement strength $\gamma$, for $\Delta = 10~\Omega,~14~\Omega \text{ and } 20~\Omega$, $\Gamma_{\rm L}=20~\Omega$ and $\Gamma_{\rm R} = 16~\Omega$. Some low $\gamma$ points in the plots for $\Delta =10~\Omega,~20~\Omega$ are omitted because of high noise.
}
    \label{fig:cross_correlation}
\end{minipage}}
\end{figure}

\subsection{Cross-correlations}

{Although trajectory-level results give valuable microscopic insight into tunneling dynamics, they are difficult to verify experimentally because resolving individual jumps requires temporal resolution comparable to the short tunneling times. For this reason we focus on the zero-frequency cross-correlation \cite{buttiker:1992, clerk2010introduction} between $I_{\rm T}$ and the detector current $I_{\rm Q}$ (Fig.~\ref{fig:cross_correlation}). This observable is experimentally convenient and can be obtained by long-time averaging of current fluctuations using standard noise-measurement techniques and therefore does not require single-shot detection. The detector current is given by 
\begin{equation}
I_{\rm Q} = \frac{e^2V}{h} [T_0 + (T_1 - T_0)(\rho_{\rm CC} + \frac{1}{\sqrt{\gamma}}\frac{dW}{dt})].
\end{equation}
}
Figure~\ref{fig:cross_correlation}(a) shows the zero frequency cross-correlation between $I_{\rm T}$ and $I_{\rm Q}$, defined as \cite{buttiker:1992, clerk2010introduction}
\begin{equation}
\begin{aligned}
\label{eq:cross_correlation}
    S_{\rm TQ}(0) = \int_0 ^{\infty}dt~\langle &\delta I_{\rm T}(t_0) \delta I_{\rm Q}(t_0 + t) + \delta I_{\rm Q}(t_0) \delta I_{\rm T}(t_0 + t) \rangle,
\end{aligned}
\end{equation}
where $t_0$ is large enough such that the evolution has become stationary (no transient dynamics as in Fig.~\ref{fig:diffusive}(b) black curve).
{ To further resolve the dynamical origin of the sign of $S_{\rm TQ}(0)$, in Appendix~\ref{app:correlation} we analyze the corresponding time-ordered current cross-correlation $C_{\rm TQ}(\tau)=\langle \delta I_{\rm T}(t_0+\tau)\delta I_{\rm Q}(t_0)\rangle$.}
We find that the currents are negatively correlated if $T_0$ is lower than $T_1$.
The Pearson coefficient, given by \cite{benesty2009pearson}
\begin{equation}
    \epsilon = \frac{S_{\rm TQ}(0)}{\sqrt{S_{\rm TT}(0) S_{\rm QQ}(0)}}
\end{equation}
and plotted in Fig.~\ref{fig:cross_correlation}(b), gives the relative strength of cross-correlations between $I_{\rm T}$ and $I_{\rm Q}$ with respect to the auto-correlations of those currents.
The auto-correlation terms are given by $S_{ii}(0) = 2 \int_0 ^{\infty}dt~\langle \delta I_{i}(t_0) \delta I_{i}(t_0 + t) \rangle$, where $i \in {\rm \{T, Q\}}$. 
{Without measurement (i.e., $\gamma=0$), we expect the cross-correlation to be zero as the TQD system and QPC are uncoupled. This is verified by the green curve with $\Delta=14\Omega$.}
For $\Delta=10\Omega,~20\Omega$, the numerical results for low $\gamma$ exhibit the expected trend but get too noisy for $\gamma\to0$ to be reliably included in the plot.
As $\gamma$ increases, the cross correlations do again approach zero for all shown values of $\Delta$ as expected, since $I_{\rm T}$ itself approaches zero in this limit.
The Pearson coefficient also shows similar behavior in Fig.~\ref{fig:cross_correlation}(b), but more importantly, the two currents are almost maximally anti-correlated ($\epsilon\approx-1$) for a significant range of $\gamma$. Since both $I_{\rm{T}}$ and $I_{\rm{Q}}$ are experimentally accessible, comparing our theoretical predictions for  $S_{\rm{TQ}}(0)$  with measured values provides a direct and effective means to validate our analysis.

\section{Quantum jump measurement}
\label{sec:jumps}

We now consider a different limit of continuous measurement, where the transmittance of the QPC is much smaller than 1 and the electrons pass through its constriction via tunneling. We further assume that the transmittance is much bigger when C is not occupied compared to when it is (because of Coulomb repulsion), i.e., $T_0 \gg T_1$. This implies that whenever an electron passes through the QPC, the occupation of the central dot jumps to zero \cite{andrew_book}. 
{ This kind of measurement results in quantum trajectories that are qualitatively different from the diffusive case in that they evolve smoothly in between jumps. These jumps are, however, stochastic in nature.} The stochastic master equation is now given by \cite{andrew_book}
\begin{align}
\hat{\rho}(t + dt) = \,\, dN(t) \frac{\hat{L}_{\gamma}  \hat{\rho}(t) \hat{L}_{\gamma} ^\dagger}{\text{Tr}[\hat{L}_{\gamma}  \hat{\rho}(t) \hat{L}_{\gamma} ^\dagger]}+ \left(1 - dN(t)\right)\hat{\rho}^{\rm nj}(t + dt),
\end{align}
where ``nj'' stands for no-jump, $\hat{L}_{\gamma} = \sqrt{\gamma}(\mathbb{1} - |C\rangle \langle C|)$, and $dN(t)$ denotes the change in number of electrons detected at the QPC, which follows a binomial distribution of values 1 and 0 with probability $\gamma (1 -\rho_{\rm CC}) dt$ and $1-\gamma (1 -\rho_{\rm CC}) dt$, respectively. In between jumps, the evolution is given by
\beq
\begin{aligned}\label{eq:jump}
\hat\rho^{\rm nj}(t+dt)=~&\hat\rho(t) - i[\hat{H}_{\text{TQD}},\hat\rho]dt+\left({\cal L}_{\Gamma_{\rm L+}}\hat\rho + {\cal L}_{\Gamma_{\rm R-}}\hat\rho+{\cal L}_\gamma\hat\rho \right)dt,
\end{aligned}
\eeq
where ${\cal L}_{\Gamma_{\rm L+}}\hat\rho$ and ${\cal L}_{\Gamma_{\rm R-}}\hat\rho$ have the same form as in the diffusive case, but for the measurement superoperator, only the dissipator part is considered
\begin{equation}
{\cal L}_\gamma\hat\rho = \gamma[\frac{1}{2}(\hat\rho \rm{|C\rangle \langle C| + |C\rangle \langle C}|\hat\rho) - \rho_{\rm CC} \hat\rho].
\end{equation}
{ 
Averaging over measurement records reproduces the same unconditional dynamics as in the diffusive case. Using
\begin{equation}
\langle dN(t)\rangle
=
{\rm Tr}\!\left[\hat{L}_\gamma^\dagger \hat{L}_\gamma \hat{\rho}\right]dt
=
\gamma(1-\rho_{\rm CC})dt,
\end{equation}
we obtain
\begin{align}
\langle \hat{\rho}(t+dt)\rangle
=&~
\gamma(1-\rho_{\rm CC})dt
\frac{\hat{L}_{\gamma}\hat{\rho}\hat{L}_{\gamma}^\dagger}
{{\rm Tr}[\hat{L}_{\gamma}\hat{\rho}\hat{L}_{\gamma}^\dagger]}
+
\left[1-\gamma(1-\rho_{\rm CC})dt\right]
\hat{\rho}^{\rm nj}(t+dt).
\end{align}
Substituting Eq.~\eqref{eq:jump} for $\hat{\rho}^{\rm nj}(t+dt)$ and retaining terms only to first order in $dt$ gives
\begin{align}
\langle \hat{\rho}(t+dt)\rangle
=
\hat{\rho}(t)
+
dt\Big(
-i[\hat{H}_{\rm TQD},\hat{\rho}]
+
{\cal L}_{\Gamma_{\rm L+}}\hat{\rho}
+
{\cal L}_{\Gamma_{\rm R-}}\hat{\rho}
+
\hat{L}_\gamma\hat{\rho}\hat{L}_\gamma^\dagger
-\frac{1}{2}
\left(
\hat{\rho}\hat{L}_\gamma^\dagger\hat{L}_\gamma
+
\hat{L}_\gamma^\dagger\hat{L}_\gamma\hat{\rho}
\right)
\Big),
\end{align}
which is precisely the unconditional master equation obtained from the diffusive measurement description.
Thus, our earlier results for average $\rho_{\rm CC}$, $I_{\rm T}$, and current cross-correlations hold for jump measurement.}

\begin{figure}[!tb]
\fbox{%
\begin{minipage}{\columnwidth}
\centering
\includegraphics[width=1\linewidth]{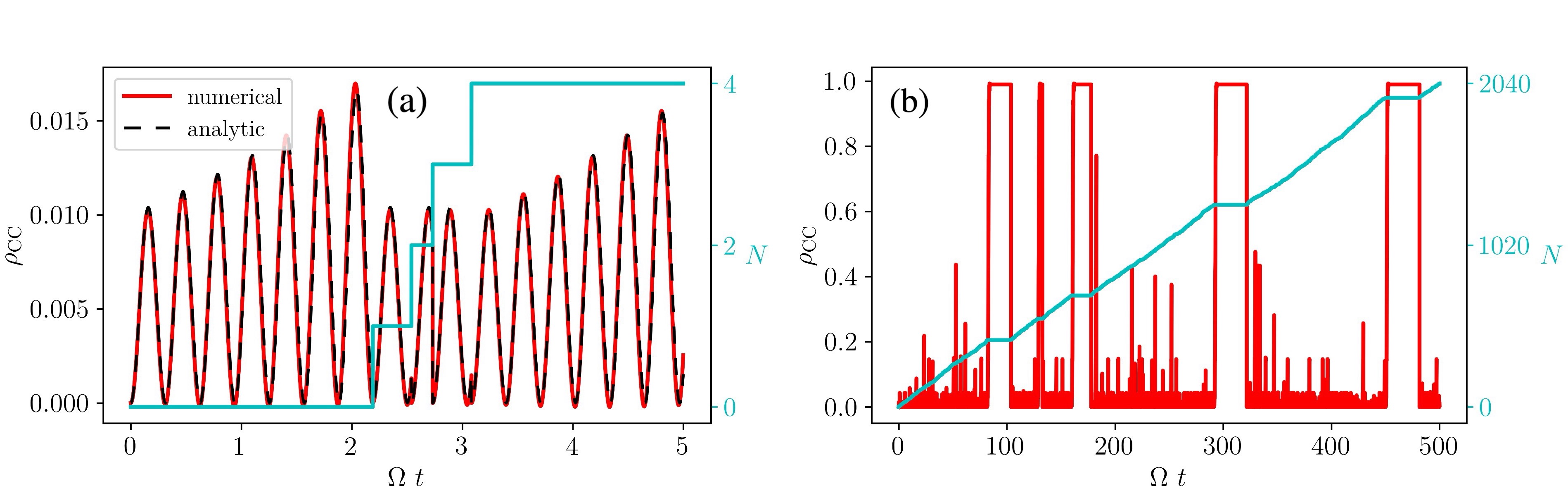}
\caption{\label{fig:jump} (a,b) Central dot occupation ($\rho_{\rm CC}$) (left axis) and number of electrons (cyan, right axis) collected at the QPC detector ($N$) as a function of time. Figure (a) contains both the numerical (red, solid) as well as an approximate analytic result (black, dashed). $\gamma$ is taken to be 0.5$\Omega$ in (a) and 5$\Omega$ in (b). Rest of the parameters are common: $\Delta=20\Omega$, $\Gamma_{\rm L}=20\Omega,~\Gamma_{\rm R}=16\Omega$, and $dt=10^{-4}\Omega^{-1}$. Initial state is taken to be the pure
state ${\rm |L\rangle \langle L|}$.
}
\end{minipage}}
\end{figure}

The central dot occupation as a function of time is shown in Fig.~\ref{fig:jump}(a), along with the total number of electrons collected at the detector as a function of time, $N(t) = \sum_{t_i=0}^{t}dN(t_i)$ (see the solid cyan curve). Although the detection of an electron in the QPC projects $\rho_{\rm{CC}}$ to zero, the no-jump evolution leads to a relatively higher occupation, on average, compared with the no-measurement scenario. 
{This is expected since the ensemble-average dynamics, as noted above, is independent of the kind of measurement done, and measurement of any strength provides an improvement in the average barrier occupation (see Fig.~\ref{fig:diffusive}(b)). 
This can be better understood with the help of the following approximate analytic expression, for the evolution of $\rho_{\rm CC}$ in-between jumps shown in Fig.~\ref{fig:jump}(a):
\beq \begin{aligned}
\rho_{\rm{CC}}(\Tilde{t}) = \frac{2\Omega^2}{\Delta^2} e^{\gamma \Tilde{t}/2}(1 -\cos(\Tilde{t}\Delta)),    
\end{aligned} \eeq
where $\Tilde{t}$ denotes the time since the last jump occurred. As time passes before the next jump, the probability of electron's presence in C (averaged over the oscillation period, $2\pi/\Delta$) increases because of the exponential rise in the amplitudes of the oscillation. Without measurement, this exponential rise is absent.}
Note that the above expression, derived in Appendix~\ref{app:analytic}, only holds when $\Omega,~\gamma \ll \Delta$ and for short periods of time after the jump. In Fig.~\ref{fig:jump}(a), we see an excellent match between this expression and the numerical result.

In Fig.~\ref{fig:jump}(b), we look at the same setup but with a higher measurement strength, which helps us sporadically catch the electron in C, with certainty, for significant intervals of time. This is analogous to stochastic diffusive trajectories often reaching unit occupation probability in C.

\section{Conclusions}
\label{sec:conclusions}

In this article, we investigate the interplay between virtual transitions and measurement back-action, proposing an experiment based on a triple quantum dot system under the continuous measurement of a highly-detuned central dot. 
We investigated the time evolution of the central dot occupation in both the diffusive and the jump quantum continuous measurement limits. Diffusive measurements result in noisy trajectories, while jump measurements result in smooth trajectories between jumps.
{ Both of these measurement limits, however, result in the same ensemble averages for the same measurement strength.
The probability of average occupation of the central dot increases significantly due to measurement back-action, and saturates at a value of 1/3 as the measurement strength grows.}
{ Although the average current through the TQD setup, $I_{\rm T} \propto \rho_{\rm RR}$, first experiences {\it measurement-assisted tunneling}, it asymptotically approaches zero with increasing measurement strength—a consequence of the quantum Zeno effect.}

We show that measurement can fundamentally alter the presence of an electron in the virtual state, offering a deeper insight into the role of observation in understanding the virtual states. The apparent paradox of observing the particle in the central dot, whether under weak or strong measurement, can be addressed by recognizing that the measurement process itself plays an active role in localizing the particle in the classically forbidden virtual state. The measurement process not only extracts information but also injects energy into the system, effectively enabling the particle to transiently occupy the virtual state. Notably, this suggests that quantum measurements could act as a thermodynamic resource, capable of powering transport processes that would otherwise be forbidden. Such a mechanism hints at novel strategies for designing measurement-driven quantum engines \cite{mohammady2017quantum,elouardefficient2018,elouardinteraction2020,bhandari2020continuous,gherardini2020stabilizing,jordanquantum2020,bresque_twoqubit_2021,mitchison2021charging,yanik2022thermodynamics,bhandari2023measurement,ferreira_transport_2024,zhang2024local,liu_maxwell_2026,erdman2025artificially,trigal_noninvasive_2026}, where information gain and energy transfer are intrinsically intertwined. The thermodynamic operations enabled by measurement-enhanced tunneling in our setup are further explored in Ref. \cite{sanchez2026making}.

\section*{Acknowledgments}
A.N.J. thanks Natalia Ares and Yuval Gefen for valuable discussions.
A.N.S., B.B., and A.N.J. acknowledge support from the John Templeton Foundation Grant ID 63209.  R.S. acknowledges funding from the Spanish Ministerio de Ciencia e Innovaci\'on via grants No. PID2022-142911NB-I00 and No. PID2024-157821NB-I00, and through the ``Mar\'{i}a de Maeztu'' Programme for Units of Excellence in R{\&}D CEX2023-001316-M.

\appendix

\section{Stochastic Master Equation}\label{app:SME}
{The stochastic master equation (SME) for the diffusive continuous-measurement limit is presented here, together with the coupled equations for the density-matrix elements. For a continuously monitored quantum dot in the diffusive limit, the conditional state evolution can be described using the standard It\^o stochastic master equation formalism \cite{andrew_book}. In the infinite-bias limit, the SME takes the form}
\beq\label{eq:SME}
\frac{d\hat\rho}{dt}=-\frac{i}{\hbar}[\hat{H}_{\text{TQD}},\hat\rho]+{\cal L}_{\rm \gamma,L+}\hat\rho + {\cal L}_{\rm \gamma,R-}\hat\rho+{\cal L}_\gamma\hat\rho,
\eeq
with the Lindblad superoperators associated with the reservoirs
\begin{align}
{\cal L}_{\rm \gamma,L+}\hat\rho &= \Gamma_{\rm L} \left[ \rho_{00}\hat{\Pi}_{\rm LL} - \frac{1}{2}(\hat{\Pi}_{00}\hat{\rho} + \hat{\rho}\hat{\Pi}_{00}) \right],\\
{\cal L}_{\rm \gamma,R-}\hat\rho &= \Gamma_{\rm R} \left[ \rho_{\rm RR}\hat{\Pi}_{00} - \frac{1}{2}(\hat{\Pi}_{\rm RR}\hat{\rho} + \hat{\rho}\hat{\Pi}_{\rm RR}) \right],
\end{align}
and with the measurement~
\beq \label{eq:Lindbladian_measurement}
{\cal L}_\gamma \hat{\rho}\,=\,\gamma \bigg[ \rho_{\rm CC}\hat{\Pi}_{\rm CC} - \frac{1}{2}(\hat{\Pi}_{\rm CC}\hat{\rho} + \hat{\rho}~\hat{\Pi}_{\rm CC}) + (\hat{\Pi}_{\rm CC}\hat\rho + \hat\rho~\hat{\Pi}_{\rm CC} - 2 \rho_{\rm CC}~ \hat\rho)\frac{dW}{dt} \bigg],
\eeq
where $\rho_{mn}$ stands for $\langle m|\rho|n\rangle$, and $\hat{\Pi}_{mn}$ stands for the projector operator $|m\rangle \langle n|$ with $m,~n \in$ \{L, C, R\}. $dW$ is the Gaussian noise associated with each measurement readout, having a zero mean and a standard deviation equal to $\sqrt{dt}$ ($dt$ being the time interval between measurements). 

Breaking down Eq.~\eqref{eq:SME} into differential equations for individual density matrix elements, we get
\begin{align}
    \dot{\rho}_{\rm LL} &= -i\Omega (\rho_{\rm CL} - \rho_{\rm LC}) + \Gamma_{\rm L}\rho_{00} - 2\sqrt{\gamma}\rho_{\rm CC}\rho_{\rm LL}\frac{dW}{dt}, \\
    \dot{\rho}_{\rm RR} &= -i\Omega (\rho_{\rm CR} - \rho_{\rm RC}) - \Gamma_{\rm R}\rho_{\rm RR} - 2\sqrt{\gamma}\rho_{\rm CC}\rho_{\rm RR}\frac{dW}{dt}, \\
    \dot{\rho}_{\rm CC} &= -i\Omega (\rho_{\rm LC} - \rho_{\rm CL} + \rho_{\rm RC} - \rho_{\rm CR}) + 2\sqrt{\gamma}\rho_{\rm CC}(1 - \rho_{\rm CC})\frac{dW}{dt}, \\
    \dot{\rho}_{\rm LC} &= -i \left( \Omega \rho_{\rm CC} - \Delta \rho_{\rm LC} - \Omega \rho_{\rm LL} - \Omega \rho_{\rm LR} \right) - \frac{\gamma}{2} \rho_{\rm LC} - 2\sqrt{\gamma}\rho_{\rm LC}(\rho_{\rm CC} - \frac{1}{2})\frac{dW}{dt}, \\
    \dot{\rho}_{\rm RC} &= -i \left( \Omega \rho_{\rm CC} - \Delta \rho_{\rm RC} - \Omega \rho_{\rm RR} - \Omega \rho_{\rm RL} \right) - \frac{\Gamma_{\rm R} + \gamma}{2} \rho_{\rm RC} - 2\sqrt{\gamma}\rho_{\rm RC}(\rho_{\rm CC} - \frac{1}{2})\frac{dW}{dt}, \\
    \dot{\rho}_{\rm LR} &= -i\Omega (\rho_{\rm CR} - \rho_{\rm LC}) - \frac{\Gamma_{\rm R}}{2} \rho_{\rm LR} - 2\sqrt{\gamma}\rho_{\rm CC}\rho_{\rm LR}\frac{dW}{dt}.
\end{align}

To obtain the results shown for the diffusive limit in the article, we solve the above equations numerically. The ensemble average evolution can be obtained by dropping the stochastic terms in these equations.

\begin{figure}[t]
\fbox{%
\begin{minipage}{\columnwidth}
\centering
\includegraphics[width=.6\linewidth]{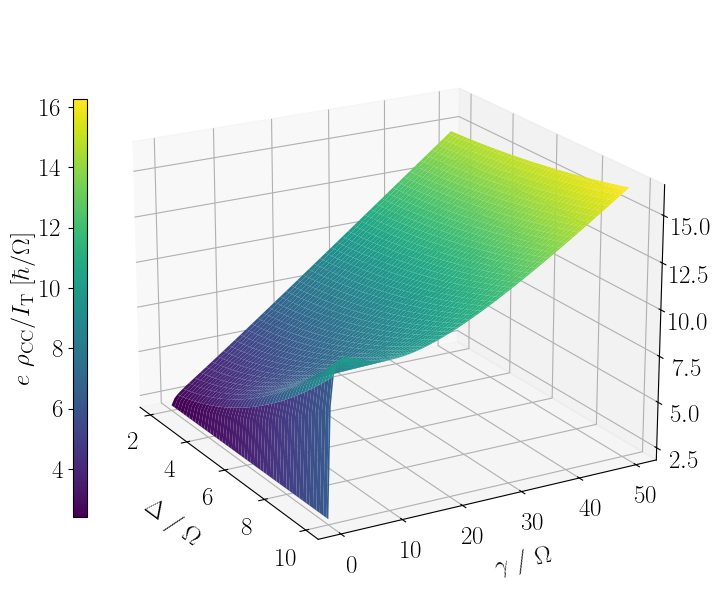}
\caption{ 
Ensemble-averaged dwell time ($e\,\rho_{\rm CC}/I_{\rm T}$) for different values of $\Delta$ and $\gamma$. Parameters: $\Gamma_{\rm L}=10~\Omega$, and $\Gamma_{\rm R}=8~\Omega$. 
    \label{fig:dwell}}
\end{minipage}}
\end{figure}

{
\section{Unmonitored Occupations}
\label{app:nomeasurement}
In the case where the detector is not connected, $\gamma=0$, we obtain simple analytic expressions for the occupations, $\rho_{ii}^{(0)}=r_i/\sum_ir_i$, with:
\begin{align}
r_0&=4\Gamma_\R\Omega^4\\
r_\L&=\Gamma_\L[\Gamma_\R^2(\Delta^2+\Omega^2)+4\Omega^4]\\
r_\C&=\Gamma_\L\Omega^2(\Gamma_\R^2+4\Omega^2)\\
r_\R&=4\Gamma_\L\Omega^4.
\end{align}
In the limit $\Delta\gg\Omega$, all occupations are suppressed except $\rho_{\L\L}$: the electron gets trapped before the barrier. Note that $\rho_{\L\L}=\rho_{\C\C}$ in the hopping regime, where $\Delta=0$.

For the parameters chosen in Fig.~\ref{fig:diffusive}, the steady-state populations are:  $\rho_{\rm LL} = \rho_{\rm CC}= 0.475$, $\rho_{\rm RR}=0.028$, and $\rho_{00}=0.022$ (when $\Delta=0$), and $\rho_{\rm LL} = 0.9885$, $\rho_{\rm CC} = 0.0104,$ $\rho_{\rm RR} = 0.0006$, and $\rho_{00} = 0.0005$ (when $\Delta=10\,\Omega$).
}

\section{Dwell Time}\label{app:dwell}

Our analysis provides us with the means to calculate the ensemble-averaged dwell time an electron spends inside the discrete tunnel barrier when the steady state has been achieved. Note that the dwell time \cite{buttiker_traversal_1982} is not conditioned upon the electron being subsequently transmitted, and thus, it is not the traversal time. For our setup, it is given by $e~\rho_{\rm CC}/|I_{\rm T}|$ ($e$ being the electron charge) where both the occupation and the current are ensemble-averaged, steady state values. Fig.~\ref{fig:dwell} shows how it varies as a function of $\Delta$ and $\gamma$. It diverges at infinitely strong measurements because the electron gets trapped inside the barrier for long periods of time whenever it is captured in this regime.

{
\section{Time-ordered Cross-correlation}\label{app:correlation}

To further characterize the interplay between the transport current through the TQD and the detector current through the QPC, we evaluate the time-ordered current cross-correlation function in the steady-state regime. We focus on the following correlation function,
\begin{equation}
C_{\rm TQ}(\tau)
=
\left\langle
\delta I_{\rm Q}(t_0)\,\delta I_{\rm T}(t_0+\tau)
\right\rangle,
\label{eq:CTQ}
\end{equation}
where $\delta I_\alpha(t)=I_\alpha(t)-\langle I_\alpha\rangle$ for $\alpha\in\{{\rm T,Q}\}$ and $\langle \cdots \rangle$ denotes an ensemble average over stochastic trajectories ($t_0$ is chosen to be high enough such that we are close to the steady-state regime).
For the diffusive measurement considered in the main text, the detector current is given by
\begin{equation}
I_{\rm Q}(t) = \frac{e^2V}{h}\bigg(T_0+\left(T_1-T_0\right)
\left[\rho_{\rm CC}(t)+\frac{1}{\sqrt{\gamma}}\frac{dW(t)}{dt}\right]\bigg),
\label{eq:IQcorr}
\end{equation}
Here $dW(t)$ is the same Wiener increment appearing in the stochastic master equation, satisfying $dW(t)^2=dt$ in the It\^o representation. The transport current is defined as
\begin{equation}
I_{\rm T}(t)=\Gamma_{\rm R}\rho_{\rm RR}(t).
\end{equation}

\begin{figure}
\fbox{%
\begin{minipage}{\columnwidth}
    \centering
    \includegraphics[width=0.7\linewidth]{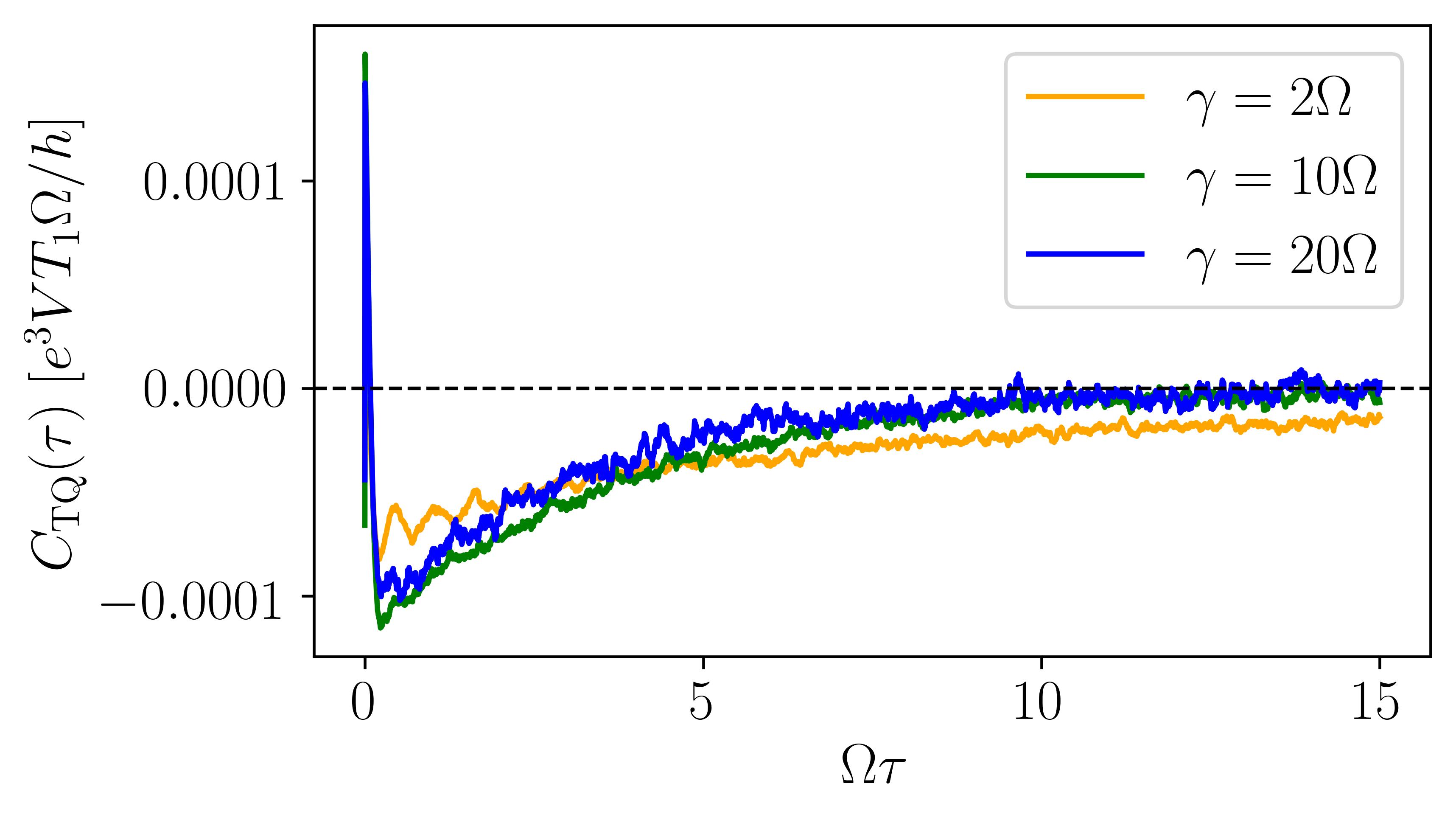}
    \caption{
Time-ordered connected cross-correlation $C_{\rm TQ}(\tau)=\langle \delta I_{\rm T}(t+\tau)\delta I_{\rm Q}(t)\rangle$ between the TQD and QPC current for three values of $\gamma$, evaluated in the diffusive measurement regime. Other parameters: $\Delta=10\,\Omega$, $\Gamma_{\rm L}=20\,\Omega$, and $\Gamma_{\rm R}=16\,\Omega$.}
    \label{fig:time_corr}
\end{minipage}}
\end{figure}

Figure~\ref{fig:time_corr} shows $C_{\rm TQ}(\tau)$ for $\Delta=\gamma = 10\,\Omega$, where only positive $\tau$ are considered to show the effects of measurement outcomes on the TQD current. There are three interesting features of the plot: equal-time negative correlation, short-time positive correlation for $\Omega\tau\ll1$, and long-time negative correlation relaxing toward zero.

The finite equal-time anticorrelation reflects that the same stochastic trajectories that transiently favor occupation of the central dot also tend to enhance occupation of the right dot, leading to positive covariance between $\rho_{CC}$ and $\rho_{\rm RR}$. Since the QPC current decreases with increasing $\rho_{\rm CC}$, these correlated fluctuations translate into a negative equal-time cross-correlation between $I_{\rm Q}$ and $I_{\rm T}$.

The short-time positive correlation occurs because a high QPC current corresponds to reduced occupation of the monitored central dot, since $T_0>T_1$. By probability conservation, such a reduction in $\rho_{\rm CC}$ can be accompanied at short times by enhanced occupation of the outer-dot subspace, including $\rho_{\rm RR}$, which directly increases the transport current $I_{\rm T}=\Gamma_{\rm R}\rho_{\rm RR}$.

The negative correlation at longer delays can be understood as a consequence of the transport sequence itself. A fluctuation that transiently increases the occupation of the central dot enhances the probability of subsequent transfer into the right dot, thereby increasing the transport current at a later time. Since the QPC current decreases when the central dot is occupied ($T_0>T_1$), this delayed population transfer leads to a negative contribution to the time-ordered cross-correlation. The eventual decay
\begin{equation}
C_{\rm TQ}(\tau\rightarrow\infty)\rightarrow0
\end{equation}
confirms loss of memory at long times. The zero-frequency cross-correlation discussed in the main text is recovered from the time-domain correlator through
\begin{equation}
S_{\rm TQ}(0)
=
\int_{-\infty}^\infty d\tau\, C_{\rm TQ}(\tau),
\end{equation}
where the negative $\tau$ contributions are also included.
}

\section{Quantum Jump Case: Analytic Expression}\label{app:analytic}
The state evolution for the quantum jump limit of measurement looks simple enough to warrant an analytic expression, when $\gamma \ll \Delta$, as shown in Fig.~4 of the main text. It follows the structure of a sinusoidal oscillation increasing in amplitude with time. Frequent jumps ensure that $\rho_{\rm CC}$ and $\rho_{\rm RR}$ remains close to zero, while $\rho_{\rm LL}$ is close to one. To derive the analytic expression, we thus assume $\rho_{\rm RR}=0$, $\rho_{\rm LL}=1$, and $\rho_{\rm CC}\ll\rho_{\rm LL}$. The coherence between the left and the right dot, $\rho_{\rm LR}$, is also negligible since $\Omega \ll \Delta$, and thus, it is taken to be 0. All of these assumptions were verified by the numerics to have a negligible effect on the evolution. The differential equations left to solve are
\begin{equation}
    \dot{\rho}_{\rm CC} = 2\Omega\Im{\rho_{\rm LC}} + \gamma\rho_{\rm CC},
\end{equation}
\begin{equation}
    \Re\dot{\rho}_{\rm LC} = -\Delta\Im{\rho_{\rm LC}} + \gamma\Re\rho_{\rm LC}/2,
\end{equation}
\begin{equation}
    \Im\dot{\rho}_{\rm LC} = \Omega + \Delta\Re{\rho_{\rm LC}} + \gamma\Im\rho_{\rm LC}/2.
\end{equation}
By substitution, this reduces to a single third order differential equation, which can be exactly solved. The analytic expression for $\rho_{\rm CC}$, in between jumps, is then given by 
\beq
\rho_{\rm CC}(t) = 2\frac{\Omega^2}{\Delta^2}e^{\gamma t/2}(1-\cos{t\Delta} ),
\eeq
under these approximations.


\bibliographystyle{iopart-num}
\bibliography{references.bib}

\end{document}